\documentclass[aps,prd,twocolumn,showpacs,superscriptaddress,nofootinbib]{revtex4}
\usepackage[english]{babel}
\usepackage{amsmath,amssymb}
\usepackage{graphicx}

\newcommand{\lp}{l_{\mathrm P}}

\newcommand{\tr}{\mbox{tr}}
\newcommand{\be}{\begin{equation}}
\newcommand{\ee}{\end{equation}}
\newcommand{\bq}{\begin{eqnarray}}
\newcommand{\eq}{\end{eqnarray}}

 \newcommand {\rhoc}{\rho_{\mathrm{c}}}
 
\newcommand{\rhom}{\rho_{M}}
\newcommand{\Pm}{p_{M}}
\newcommand*{\R}{{\mathbb R}}
\newcommand{\RBohr}{\R_{\mathrm{Bohr}}}

\newcommand*{\C}{{\mathcal C}}
\newcommand*{\Ham}{{\mathcal H}}
\newcommand*{\Or}{{\mathcal O}}
\newcommand{\E}{E}

\newcommand{\ket}[1]{\left| #1 \right\rangle}

\newcommand{\braket}[2]{\langle #1 \,|\, #2 \rangle }
\newcommand{\braketfull}[3]{\langle #1 \,|\, #2 \,|\, #3 \rangle }
\newcommand{\qz}{{}^o\!q}
\newcommand{\ez}{{}^o\!e}
\newcommand{\wz}{{}^o\!\omega}
\newcommand{\sqz}{\sqrt{\qz}}

\newcommand{\half}{\frac{1}{2}}
\newcommand{\sgn}{\text{sgn}}

\newcommand{\Pp}{P_{\phi}}

\newcommand{\g}{\gamma}
\newcommand{\G}{\Gamma}

\newcommand{\kap}{\kappa}

\newcommand{\tp}{\widetilde{p}}
\newcommand{\tc}{\widetilde{c}}

\newcommand{\wh}[1]{\widehat{#1}}

\newcommand{\Om}{V_0}

\newcommand{\m}{\mu}

\newcommand{\kn}{\ket{v}}

\newcommand{\ra}{\rightarrow}

\newcommand{\non}{\nonumber}

\newcommand{\mb}{\overline{\m}}

\newcommand{\CGR}{\C_{GR}}

\newcommand{\intt}{\int\!\!dt}

\begin{document}

\title{Loop quantum cosmology and the k = - 1 RW model}
\author{Kevin Vandersloot}
\email{Kevin.Vandersloot@port.ac.uk}
\affiliation{Institute for Gravitational Physics and
Geometry,\\ The Pennsylvania State University,\\ University Park, PA
16802, USA}
\affiliation{Institute for Cosmology and Gravitation,\\ University of Portsmouth,\\
Portsmouth, PO1 2EG, UK}

\begin{abstract}
The loop quantization of the negatively curved $k=-1$ RW model poses several 
technical challenges.
We show that
the issues can be overcome and a successful quantization is possible that 
extends the results of the $k=0,+1$ models in a natural fashion.
We discuss the resulting
dynamics and show that for a universe consisting of a massless scalar field,
a bounce is predicted in the backward evolution in accordance with the
results of the  $k=0,+1$
models. We also show that the model predicts
a vacuum repulsion in the high curvature regime that would lead to a bounce
even for matter with vanishing energy density. We finally comment on the
inverse volume modifications of loop quantum cosmology and show that,
 as in the $k=0$ model, the modifications  depend sensitively
on the introduction of a length scale  which a priori is independent of
the curvature scale or a matter energy scale. 
\end{abstract}

\pacs{98.80.Qc, 04.60.Pp, 04.60.Kz, 03.65.Sq}

\maketitle

\section{Introduction}

One of the major cosmological parameters yet to be determined precisely pertains to
the spatial curvature of the universe. While current observations indicate that the universe
is very nearly flat, they do not yet provide irrefutable evidence as to whether on very large
scales the curvature is positive, negative, or exactly zero (the $k=+1, -1, 0$ Robertson-Walker (RW) models
respectively). The current observations
merely provide evidence of a prediction of inflation; namely, the curvature scale is sufficiently
large such as to appear very nearly flat to an observer. Such a feature is explained by inflation
through a period of accelerated expansion in the early universe that inflates the curvature
scale to very large values. Thus most work on structure formation has assumed
that the universe is exactly flat with $k=0$ which is a good approximation
for the post inflationary epoch. However, it is the period of the early universe where
the curvature can play an important role and thus should not be neglected.

It is also the high energy regime of the early universe where quantum gravity is expected
to be a requirement for a complete description. While no complete and fully accepted quantum
theory of gravity exists, a leading candidate exists which is known as loop quantum gravity (LQG)
\cite{ash10,book, Thiemann:2001yy}. The application of LQG techniques to the cosmological setting,
loop quantum cosmology (LQC), 
has so far been restricted to the $k=0,+1$ models (see \cite{Bojowald:2006da} for a review). One of the major successes of the models
of LQC so far is the resolution of the classical singularity predicted in the $k=0,+1$ models \cite{Bojowald:2001xe}
which can result in a repulsive gravitational force at high energies that leads to a big-bounce
of the universe \cite{Ashtekar:2006rx, Ashtekar:2006uz, Ashtekar:BBII, Newkplus1}. 
Thus an open question remains as to whether these results hold in the negatively curved $k=-1$
model and whether a loop quantization even exists.

The  $k=-1$ model has not been constructed in LQC owing
to technical issues that inhibit a successful quantization. The $k=-1$ model
can be derived as the isotropic limit of the homogeneous Bianchi V model which
lacks a correct Hamiltonian framework \cite{Maccallum:1972vb}. The Hamiltonian
framework is essential to the canonical quantization scheme of both LQG and LQC
and thus this failure presents a roadblock to quantization. Notwithstanding this issue, as we shall
show the $k=-1$ model also leads to subtle features in the choice of dynamical
variables in LQC that require careful attention when attempting a quantization.

In this paper we will show that these issues can be successfully overcome leading
to a loop quantization of the model. We will show that the Hamiltonian framework
can be constructed specifically for the isotropic Bianchi V model and that
the theory can be quantized
incorporating techniques similar to those used in the loop quantization of spherically symmetric models.
The resulting quantum theory is  in a form that is similar to the $k=0,+1$ LQC models
and thus shares many of the same features. We show directly that the model
predicts a big-bounce in the backward evolution of the universe sourced
by a massless scalar field. We describe this behavior in terms of
an effective Friedmann equation that is quadratic in the matter energy density.
Furthermore the effective Friedmann equation predicts a vacuum repulsion
in the Planckian curvature regime, whereby a bounce would
be triggered even with vanishing matter density. Finally, we comment
on the inverse volume effects predicted by LQC and show that they
are dependent on the introduction of a scale into the model which
is not determined from the curvature scale or any matter energy scale.
We discuss the phenomenological implications of this.

\section{Classical Framework}

We begin  with the classical framework that will
form the basis of the loop quantization for the $k=-1$ model.
Loop quantum gravity (and hence loop quantum cosmology) is based
on a Hamiltonian framework using connection-triad variables as
the gravitational field variables. The goal of this section is to consider
the connection-triad variables which are invariant under the symmetries
of the Bianchi V group (which leads to the $k=-1$ model), and then construct
the Hamiltonian in terms of the reduced variables, and finally show that the equations
of motion derived from the Hamiltonian give back the usual cosmological equations
of motion for the open model.

The starting point for homogeneous cosmological model we consider
are the Bianchi models. The homogeneous metric 
is given by
\bq
    ds^2 = - N(t)^2 \;dt^2 + \alpha_{ij}(t)\;\, \wz^i_a  \, \wz^j_b \; dx^a dx^b
\eq
where $\alpha_{ij}(t)$ are the dynamical components of the metric,
$N(t)$ is known as the lapse and represents the rescaling freedom of the
time coordinate, 
and $\wz_a^i$ are a basis of left-invariant one-forms determined by
the group structure of the Bianchi model being considered.
The left-invariant one-forms satisfy
\be \label{MC}
    d\,\wz^i = - \half  {C^i}_{jk} \;\wz^j \wedge \wz^k.
\ee
where ${C^i}_{jk}$ are the structure constants of
the isometry group and thus characterize the Bianchi model.
For the open $k=-1$ model, we consider the Bianchi V model
with structure constants that can be taken of the form
\bq \label{cijk}
    {C^i}_{jk} = \delta^i_k \delta_{j 1} - \delta^i_j \delta_{k 1} \,.
\eq
The structure constants satisfy ${C^i}_{ij} \ne 0$ which
in the language of \cite{Ellis:1968vb} implies that the Bianchi V model is class B. This fact
will be important in what we consider later.

In a particular choice of coordinates, 
equations (\ref{MC}) can be solved explicitly to give the left-invariant
one-forms as
\bq
    \wz^1 &=&   dx  \\
    \wz^2 &=&  e^{-x} dy   \\
    \wz^3 &=&  e^{-x} dz \,.
\eq
where   the  coordinates $x,y,z $ are valued on the real line representative of the fact
that we are considering the spatially non-compact k=-1 model with topology
homeomorphic to $\R^3$. Thus we have not chosen a particular
compactification of the k=-1 model and work in the usual model
with infinite spatial extent. If we consider the isotropic limit of this
Bianchi model with $\alpha_{ij}(t) = a^2(t) \delta_{ij}$ with $a(t)$ representing
the scale factor, and fix the lapse to be equal to one, then the metric
\bq \label{cmetric}
	ds^2 &=& - dt^2 + a^2 \, \delta_{ij}\; \wz^i_a \, \wz^j_b \; dx^a dx^b \non \\
	&=& -dt^2 + a^2 ( dx^2 + e^{-2x} dy^2 + e^{-2x} dz^2)
\eq
can be shown to have constant negative spatial curvature and 
hence corresponds to the open $k=-1$ model.
The usual hyperbolic metric in hyperbolic coordinates
$ds^2 = -dt^2 + a^2(d\psi^2 + \sinh^2\psi \big( d\theta^2 + \sin^2\theta d\phi^2))$ can be recovered with the following transformation
\bq
    x &=& - \ln(\cosh\psi -\sinh\psi \cos\theta) \non \\
    y &=& \frac{\sin\theta \cos\phi}{\coth\psi - \cos\theta} \non \\
    z &=& \frac{\sin\theta \sin\phi}{\coth\psi - \cos\theta} \non \,.
\eq
With the form of the metric (\ref{cmetric}), Einstein's equations
lead to a set of differential equations satisfied by the scale factor $a(t)$ given
by the Friedmann equation
\be \label{Fried}
    \Big(\frac{\dot{a}}{a}\Big)^2 = \frac{\kappa}{3} \rho_M + \frac{1}{a^2}
\ee
and the acceleration equation
\bq \label{accel}
    \frac{\ddot{a}}{a} = -\frac{\kap}{6}\Big(\rhom + 3 \Pm\Big)
\eq
with $\rhom$ and $\Pm$ being the matter density and pressure respectively.
Here $\kappa=8\pi G$ with $G$ being Newton's constant. 

In addition to the left-invariant one-forms, for what follows we
will also need a basis of vector fields $\ez^a_i$ which also are left-invariant.
The left-invariant
vector fields have commutators which provide a representation of the Lie
algebra under consideration
\bq \label{coms}
    [\ez_i, \ez_j] = {C^k}_{ij}\; \ez_k
\eq
and are also dual to $\wz_a^i$ thus satisfying 
\bq
	\ez^a_i \; \wz_a^j = \delta^j_i \,.
\eq
In the chosen coordinates for the Bianchi V model, $\ez^a_i$ are
given explicitly as
\bq
    \ez_1 = \partial_x, \;\;\;\;\;
    \ez_2 = e^{x}\, \partial_y, \;\;\;\;\;
    \ez_3 = e^x \,\partial_z
\eq
whence it is simple to show satisfy equation (\ref{coms}).

The classical framework of loop quantum cosmology (LQC) diverges from the standard
framework in two important ways. The first is that the equations of motion
are derived from a Hamiltonian framework which allows for a canonical quantization
of the theory. Secondly, the variables that form the basis for quantization are not
the usual metric ones (i.e., the scale factor). This framework
follows directly from that used in the full theory of loop quantum gravity
and it is these changes that allow for a rigorous quantization of gravity. The canonical
set of variables consists of an orthonormal triad $\E^a_i$ (of density weight one) which encodes the information
of spatial geometry, and an su(2) valued connection $A_a^i$ which is canonically conjugate
to $\E^a_i$. The starting point of LQC is to reduce these variables to the symmetry
of the cosmological model. We can use the basis provided by the left-invariant
one-forms and vector fields to accomplish this.

Starting with the triad $\E^a_i$, we expand using the basis vector-fields as
\bq \label{EBV}
	\E^a_i =  \sqz\; \tp(t) \; \ez^a_i
\eq
where $\tp(t)$ represents the dynamical component of the triad. 
The factor $\sqz = e^{-2x}$ is a density weight provided by the 
hyperbolic metric $\qz_{ab} = \wz_a^i \,\wz_b^i$ which 
gives the triad $\E^a_i$ its density weight.
$\E^a_i$ encodes the spatial geometry in a specific fashion being that it is
related to the spatial three-metric $q_{ab}$ through
\bq\label{Etom}
	\E^a_i \E^{b i} = |q|\, q^{ab} \,.
\eq
Using this relation, we find that $\tp$ is related to the scale factor as
\bq
	|\tp| = a^2
\eq 
where the absolute value indicates that we are allowing $\tp$ to take on positive and
negative values in contrast to the scale factor which is usually assumed to be strictly non-negative. A change in sign of $\tp$ corresponds to a change in orientation of
the triad $E^a_i$ leaving the metric $q_{ab}$ invariant.

The first non-triviality of the $k=-1$ arises when we consider a symmetric
connection $A_a^i$. From the $k=0, +1$ models, we expect that an isotropic
connection can be decomposed using the left-invariant one-forms
as $A_a^i = \tc(t)\, \wz_a^i$ \cite{Bojowald:2002gz,Bojowald:2003mc} with
$\tc$ being the {\em only} dynamical component. In this form, the connection is diagonal
in the basis of  left-invariant one-forms.
However, this form must be consistent with the fact
that  on the half-shell (after solving Hamilton's equations for $\dot{E}^a_i$), $A_a^i$ 
is determined from the dynamics of the spatial metric as
\bq
	A_a^i = \g K_a^i + \G_a^i
\eq
where $K_a^i$ is the extrinsic curvature, $\g$ is known as the Barbero-Immirzi parameter (a real valued ambiguity parameter of loop quantum gravity),
and $\G_a^i$ is the spin-connection.
Upon symmetry reduction, the extrinsic curvature can be shown to be of diagonal 
form\footnote{The $\sgn(p)$ arises
because the extrinsic curvature one-form carries the signature
of the triad which is evident from the definition $K_a^i = e^{a i} K_{ab}$ where
$K_{ab}$ is the usual extrinsic curvature of the ADM formulation which does not
carry information about the orientation.}
$K_a^i = \sgn(\tp) \dot{a}\, \wz_a^i$ 
 which is consistent with the connection
being diagonal.
However, this is not the case with  the spin connection $\G_a^i$.
The formula for the spin connection is given by
\bq
	\label{spinconnection}
	\Gamma^i_a = - \half \epsilon^{ijk} e^b_j\Big( \partial_a e_b^k - \partial_b e_a^k
		+e^c_k e_a^l \partial_c e_b^m \delta_{lm} \Big) \,.
\eq
where $e_a^i$ is the physical triad  satisfying
\bq
	e_a^i e_b^i = q_{ab} \,.
\eq
The physical triad $e_a^i$ is related to $\E^a_i$ through
\bq
	  e^a_i  = \frac{1}{\sqrt{|q|}} \; \E^a_i \,.
\eq
Using the symmetric form of $\E^a_i$ (\ref{EBV}) and evaluating
 (\ref{spinconnection}), one finds that the spin connection is
given by
\bq
	\G_a^i = \G^i_j \, \wz_a^j
\eq
with
\be
	\Gamma^i_j = \left( \begin{array}{ccc} 0&0&0\\0&0&-1\\0&1&0 \end{array} \right) 
\ee
whence it is clear that the spin connection is non-diagonal, and the assumption that the connection
is diagonal is not consistent. We must therefore take the connection to be of non-diagonal
form
\bq \label{Atwo}
	A_a^i &=& A^i_j(t) \, \wz_a^j \non \\
	A^i_j &=& \left( \begin{array}{ccc} \tc(t) &0&0\\0&\tc(t) &-\tc_2(t)\\0&\tc_2(t)&\tc(t)  \end{array} \right) \,.
\eq
In this form, the connection has two dynamical components $\tc$ and $\tc_2$, where on the half-shell
$\tc = \sgn(\tp) \g \dot{a}$ is determined from the extrinsic curvature and $\tc_2 = 1$. 
This is in contrast to the $k=0,+1$ models where the connection can safely be
assumed to be diagonal and  only has one dynamical component.

With the symmetry reduced connection-triad variables,
the next step is to show that the Hamiltonian formulation leads to the correct classical equations of motion. Yet here another problem
arises. The Bianchi V model is of class B type where, as first shown in
\cite{Maccallum:1972vb}, the ADM Hamiltonian formulation in the general
homogeneous case fails. The main issue is that the equations of motion derived from the symmetry reduced Hamiltonian
do not agree with Einsteins' equations after symmetry reduction. In other words,
the symmetry reduction and Hamiltonian formulation do not commute in the class B models.
While this may seem a fatal issue for the k=-1 model, it was shown in
\cite{Maccallum:1972vb} that the Hamiltonian formulation does not
fail for the isotropic limit of the Bianchi V model which is precisely the case
we are interested. This  failure does hinder the extension of the results
presented here to the anisotropic Bianchi V model, but we will now show explicitly that the Hamiltonian
formulation of the isotropic model using the connection-triad variables leads to the
correct equations of motion. Another avenue worth exploration is whether the analysis
of \cite{Maccallum:1972vb} holds in general for the Hamiltonian formulation based on
the connection-triad variables used in loop quantum gravity which is
manifestly different than the ADM Hamiltonian formulation and thus may not
suffer from the same issues. We do not attempt to address this possibility here.

Therefore, our aim now is to plug in the symmetry reduced connection (\ref{Atwo}) and
triad (\ref{EBV}) into the Hamiltonian of the full theory and show that we get back
the correct equations of motion (\ref{Fried}, \ref{accel}). The action
written in terms of the connection-triad variables\footnote{
The action can also be derived from a Legendre transform
of the covariant Holst action  written in terms of a
four dimensional so(3,1) connection and a co-tetrad\cite{Holst:1995pc}.} is given as \cite{ash10}
\bq \label{fullaction}
	S_{GR}[\E, A, \lambda^i, N^a, N]  = \intt \int \!\!d^3 x \frac{1}{\kappa \g} \E^a_i
		\mathcal{L}_t A_a^i  \non \\
	- \left[ \lambda^i G_i + N^a \C_a + N \CGR \right] 
\eq
whence the Hamiltonian is a sum of constraints: $G_i$ is the Gauss constraint, $\C_a$ is the
diffeomorphism constraint, and $\CGR$ is the Hamiltonian constraint. The parameters
$\lambda^i, N^a, N$ are Lagrange multipliers which enforce the vanishing of the constraints.
The first term of the action indicates that the connection and triad are
canonically conjugate with Poisson brackets
\bq
	\{ A_a^i(x), \E^b_j(y)\} = \kappa \g \,\delta^i_j \, \delta_a^b \,\delta(x-y) \,.
\eq
Hamilton's equations
for the connection $A_a^i$ and triad $\E^b_j$ can then be shown to be equivalent to
Einstein's equations.

When inserting the symmetry reduced connection and triad into the action, the first issue
we face is that the spatial integration in 
the action diverges since we are considering the non-compact $k=-1$ model. This
same issue arises in the non-compact $k=0$ model and would arise in any
cosmological quantization  scheme based on a Hamiltonian or action framework.
To overcome this, we choose the follow the technique used in the $k=0$ model
for LQC \cite{Ashtekar:2003hd}; namely, we restrict the spatial integration to a finite sized fiducial cell with
a fixed background volume 
\bq
	\Om = \int\!\! d^3x\, \sqz \;.
\eq 
Note that the extent of the fiducial cell is fixed on the manifold or
in other words has fixed comoving coordinates. Thus, as the universe
expands for instance, so would physical size of the fiducial cell. The choice in
the fiducial cell remains a quantum ambiguity and we will be interested in
determining whether  the resulting quantum theory makes predictions
dependent on $\Om$. As we now show, the choice in fiducial cell
has no effect classically, but that is not true in the quantum case which we will discuss
later.

Now with the understanding that we are limiting the spatial integrations in the action
to the fiducial cell we can insert 
the symmetry reduced connection and triad  (\ref{Atwo}, \ref{EBV}).
The canonical term is given by
\bq
	\intt \int \!\!d^3 x \,\frac{1}{\kappa \g}\, \E^a_i\,
		\mathcal{L}_t A_a^i = \intt\; \frac{3 \Om}{\kappa\g}\, \tp\; \dot{\tc}
\eq
which indicates that $\tc$ and $\tp$ are canonically conjugate with Poisson brackets
\be \label{pbrak}
	\{ \tc, \tp \} = \frac{\kappa \g}{3 \Om} \,.
\ee
The Gauss constraint in terms of the reduced variables is given by
\bq
	G_i \equiv \partial_a \E^a_i + {\epsilon_{ij}}^{k} A_a^j \E^a_k
	= \frac{2 \Om}{\kap \g}\, \tp \;(\tc_2 - 1)\, \delta_{i 1}
\eq
and thus is non-vanishing. This is in contrast to the $k=0,+1$ models
where the Gauss constraint vanishes indicative of the fact that a complete
gauge fixing of the Gauss constraint was  performed in those models. 
This suggests that we should gauge fix the Gauss constraint by
setting $\tc_2$ to be identically equal to one. With this, the Gauss constraint
vanishes and additionally the diffeomorphism constraint $C_a$ can be shown
to vanish. 
 With this gauge fixing
the connection is now of the form
\bq\label{Afinal}
	A^i_j &=& \left( \begin{array}{ccc} \tc &0&0\\0&\tc &-1\\0&1&\tc  \end{array} \right) 
\eq
and we are now left with two dynamical phase space variables $\tp$ and $\tc$
and one surviving constraint, the Hamiltonian constraint.
 This is exactly the situation in the $k=0,+1$ models.

The dynamics of the model is now entirely encoded in the Hamiltonian constraint
which is given by
\bq
	\CGR = -  \frac{6 \Om}{\g^2}  \sqrt{|\tp|} \,\Big(\tc^2 -  \g^2\Big) 
\eq
and  the entire gravitational action becomes\footnote{The extra
factor of $2\kappa$ appearing below the lapse $N$ appears because
the Hamiltonian constraint used in previous works of LQC differs from
the  Hamiltonian constraint in the full theory given in \cite{ash10} by the factor
of $2 \kappa$. A constant factor multiplying the Hamiltonian constraint
does not affect any physical results.}
\bq
	S_{GR}[\tp, \tc, N]  &=& \intt \, \frac{3 \Om}{\kappa \g}\,\tp \;
		\dot{\tc}  \non \\
	&&- \frac{N}{2 \kap} \left[ -  \frac{6 \Om}{\g^2}  \sqrt{|\tp|} \,\Big(\tc^2 -  \g^2\Big) \right] 
\eq
whence the total Hamiltonian including matter is given by
\bq \label{Ham}
	\Ham &=& -  \frac{3  \Om N}{\kappa \g^2}  \sqrt{|\tp|}\, \Big(\tc^2 -  \g^2\Big)
	+ \Ham_M
\eq
with  $\Ham_M$ denoting the matter Hamiltonian. 

With the Hamiltonian and Poisson structure we can now derive the
classical equations of motion. We first have Hamilton's equations
$\dot{x} = \{x, \Ham\}$ for any phase-space variable $x$, 
and further the Hamiltonian itself must vanish since it is proportional
to the Hamiltonian constraint.
 Starting with Hamilton's equations
for $\tp$ we find
\bq
	\dot{\tp} = \{\tp, \Ham \} &=& - \frac{\kappa \g}{3 \Om}  \, \frac{\partial\Ham}{\partial\tc} 
	\non \\
	&=& \frac{2 \sqrt{|\tp|}}{ \g}   \tc 
\eq
where for the equations of motion we have fixed  the lapse $N=1$.
Notice that the factors of $\Om$ cancel appearing both in the numerator of
the Hamiltonian (\ref{Ham}) and in the denominator of the Poisson brackets (\ref{pbrak}).
Furthermore, 
we have assumed that the matter Hamiltonian only couples to the spatial geometry
i.e., is only a function of $\tp$ and not $\tc$. This assumption is true for scalar fields
and perfect fluids, though is not true for fermions for instance which we do not consider
(see \cite{Perez:15} for discussions for the inclusion of fermions in LQG with physical effects
dependent on the Barbero-Immirzi parameter $\gamma$).
Using $|\tp| = a^2$ we can write the left-hand side of the Friedmann
equation as
\bq
	H^2 = \Big(\frac{\dot{a}}{a}\Big)^2 = \frac{\tc^2}{\gamma^2 |\tp| } \,.
\eq
Now we can use the vanishing of the constraint to relate the right-hand side
to the matter density. Using $\Ham=0$ we find
\bq
	H^2 &=& \frac{\kap}{3} \frac{\Ham_M}{\Om \tp^{3/2}} + \frac{1}{|\tp|} \non \\
	&=& \frac{\kap}{3} \rhom + \frac{1}{a^2}
\eq
where we have used $\Ham_M/(\Om \tp^{3/2}) = \rhom$. Thus the reduced Hamiltonian
gives back the correct Friedmann equation. Similarly the acceleration equation
can be derived by considering Hamilton's equation for $\dot{\tc}$ once the matter
Hamiltonian is explicitly specified.

This derivation demonstrates explicitly that the Hamiltonian framework presented here
leads to Einsteins equations for the open $k=-1$ model. This also demonstrates
that the equations of motion classically are insensitive to the choice in fiducial
cell $\Om$ which was introduced to regulate the divergent spatial integrals
in the action and resulting Hamiltonian. Furthermore, the Hamiltonian is  similar
in form to the $k=0,+1$ models and thus is indicative that a successful
loop quantization is possible.

Before turning to the quantization, we would like to make a closer
 connection to the LQC work of the $k=0, +1$ models. There we can
define untilded variables by rescaling $\tp$ and $\tc$ by a factor
dependent on the fiducial cell as 
\bq \label{untild}
	\begin{array}{rcl}
	p &\equiv& \Om^{2/3}\, \tp  \\
	c &\equiv& \Om^{1/3} \,\tc
	\end{array}
\eq
In terms of the untilded variables, the Poisson bracket is now independent of $\Om$
\be
	\{ c, p \} = \frac{\kappa \g}{3 }
\ee
and the Hamiltonian constraint becomes
\be \label{Cuntild}
	\CGR = - \frac{3 }{\kappa \g^2} \sqrt{|p|}  \Big(c^2 - \Om^{2/3} \g^2 \Big) \,.
\ee
The relation between the rescaled triad and the scale factor is given by
\bq \label{pofa}
	|p| = \Om^{2/3} a^2  \label{ptoaV}
\eq
as well as the half-shell relation
\bq
	c = \sgn(p) \Om^{1/3} \g \dot{a}\label{ctoaV} \,.
\eq 
We will use the Hamiltonian based on the untilded variables as the starting
point for quantization. Notice that $\Om$ appears explicitly in
the Hamiltonian constraint (\ref{Cuntild}) which is in contrast
to the $k=0$ model where $\Om$ drops out of the constraint.

It will be important in the interpretation of the quantum theory to
understand the physical meaning of the variable $p$. From
the relation (\ref{pofa}), we find 
\bq \label{Vcell}
	|p|^{3/2} = \Om a^3 = V_{\text {cell}}
\eq
which one recognizes as representing the {\em physical} volume
of the fiducial cell. Note that in order to physically measure
the value of $p$, one would need prescribe the size of the fiducial
cell. For instance, if today the fiducial cell is taken to have Planckian physical
volume: $V {\text {cell}} = \lp^3$, then $p$ is similarly Planckian:
$p = \lp^2$. This can be so despite the fact that, assuming we live in
an open $k=-1$ universe, the value of the scale factor $a$ 
is astronomically large. Thus there is no direct correlation between
the value of the scale factor $a$, and the value of $p$. Again the
value of $p$ is highly dependent on the size of the fiducial cell.

Let us conclude this section with a further note about the fiducial cell. Since
we will be interested in whether the quantum predictions are sensitive to
this choice, we would like to know how the classical variables transform
under a change in its size. For instance, let us consider that
the fiducial cell is resized as
\bq
	\Om \ra \Om' = \xi^3 \Om
\eq
then from their definition (\ref{untild}), the untilded variables
transform as
\bq
	p \ra p' &=& \xi^2 p  \label{pscale} \\
	c \ra c' &=& \xi c \label{cscale} \,.
\eq
Note that the scale factor $a$ (and therefore $\tp$ and $\tc$) do not make
reference to the fiducial cell and therefore do not rescale under this change. 
The scaling of $p$ can be understood from (\ref{Vcell}) by noting that
the value of $p$ is determined from the physical volume of the cell, and
thus if the cell is enlarged, we expect the value of $p$ to be larger.
The untilded variables therefore do not classically have direct physical meaning as they
can be freely rescaled under this transformation. What is physical
classically are {\em changes} in $p$ and $c$ where for instance
the Hubble rate  $H = \frac{1}{2}\big(\frac{\dot{p}}{p}\big)$
is invariant under a resizing of the fiducial cell.

\section{Quantization}
With the classical framework completed, we can now turn to the loop
quantization of the model. To achieve this in the canonical quantization
scheme involves the following steps. First one chooses a set of basic variables
and finds a quantum representation of their algebra in order to construct
what is known as the kinematical Hilbert space. The next step is to construct
an operator corresponding to the Hamiltonian constraint that is self-adjoint in
the kinematical Hilbert space. Lastly, the physical Hilbert space consists of
wavefunctions that are annihilated by the constraint operator and
that have finite norm in a suitable physical inner product (which typically is not
equivalent to the kinematical inner product). One then interprets the theory
by evaluating expectation values of observables on physical wavefunctions.
By considering the $k=-1$ model sourced with a massless scalar field, this
program can be carried out to completion. The construction of
the $k=-1$ model presented here follows closely that of the $k=0$ model presented
in \cite{Ashtekar:BBII}, and thus we will omit many of the technical details and refer the
reader to that article for a complete description.

\subsection{Kinematical Hilbert Space}

To construct the kinematical Hilbert space we first must consider the elementary
variables that will form the basis for quantization. From the full theory
of LQG, one does not take the bare connection $A_a^i$ and triad $E^a_i$
as the basic variables. Rather, in the case of the connection,
one integrates $A_a^i$ along
edges and then exponentiates the quantity leading to  a holonomy.
The holonomy variables are then taken as the basic configuration variables.
The momentum variables are fluxes which are constructed by integrating of the triad over a two-surface.
 In the cosmological setting,  fluxes
are simply proportional to $p$ which therefore forms an elementary variable.
On the other hand, the holonomies amount to exponentials of the connection
$c$ and it is this fact that becomes the departure point of LQC from
previous versions  quantum cosmology based on a Schrodinger
type quantization of the Hamiltonian.

Thus let us consider the holonomies in detail.
In the $k=0$ model they consist
of integrating the connection along edges generated by the left-invariant
vector fields and assume the form
$h_i = \cos\big( \frac{\mb c}{2 }\big) + 2 \sin\big( \frac{\mb c}{2 }\big)
				\tau_i $
where $\mb$ is equal to the fiducial  length of the edge divided by
$\Om^{1/3}$, and
$\tau_i$ are the generators of SU(2) satisfying $\{\tau_i, \tau_j\} = {\epsilon_{ij}}^{k} \tau_k$.
With holonomies of this form, the algebra generated is that of the almost periodic functions
(which look like exponentials of the connection $e^{i \mb c}$) and the kinematical Hilbert space assumes a simple form\cite{Ashtekar:2003hd}. However, when we consider
holonomies of the connection in the $k=-1$ model considered here, they take on a more
complicated form where for instance the holonomy along the edge generated
by $\ez^a_2$ is given by
\bq \label{hA}
	&h_2(\mb)& = \cos\frac{\mb\sqrt{c^2\!+\!\Om^{2/3}}}{2} \non \\
	\!\!\!\!\!\!&+&\!\!\!\!\!\!\!\! \frac{2 \left[ c \tau_2 - \Om^{1/3} \tau_3 \right] }{\sqrt{c^2\!+\!\Om^{2/3}}}\sin\frac{\mb\sqrt{c^2\!+\!\Om^{2/3}}}{2}		
\eq
and thus the algebra generated is no longer simply that of the almost
periodic functions. Finding a representation of the algebra would be difficult.

However, we can exploit a technique used in the loop quantization of other
models such as the spherically symmetric models of LQG \cite{Bojowald:2005cb} as well
as the quantization of the Schwarzschild horizon interior \cite{Ashtekar:2005qt}. The complicated
form of the holonomies of the connection arises because of the non-diagonal form
of the connection (\ref{Afinal}). If we consider instead holonomies of the connection
minus the spin-connection (essentially holonomies of the extrinsic curvature) as
done in \cite{Bojowald:2005cb, Ashtekar:2005qt}, then the holonomies are
of a form equivalent to the $k=0,+1$ models
\bq \label{holon}
	h_i = \cos\Big( \frac{\mb c}{2 }\Big) + 2 \sin\Big( \frac{\mb c}{2 }\Big)
				\tau_i \,.
\eq
where again $c$ refers to the diagonal component of the connection
in (\ref{Afinal}). In the full theory, the extrinsic curvature is not a connection and hence
its holonomies are not defined. However, in the reduced setting we
have performed a complete SU(2) gauge fixing to arrive at symmetric connections
and thus it is possible to  regard the extrinsic curvature as a connection. The
resulting quantization will be a slight departure from that predicted by the full
theory and thus care must be taken when interpreting the results. In  
section \ref{Discussion}, we will comment on the regime where we expect the differences to occur.

Additionally we shall follow the prescription of \cite{Ashtekar:BBII} leading
to improved dynamics for LQC. Namely, in contrast to the original
literature of LQC, we assume that the parameter $\mb$ appearing in
the holonomies is a function of $p$ and not a constant. The motivation for
this can be seen as twofold. First let us consider the issue of the fiducial cell
dependence. The quantity $\mb c$ appears in the holonomies (\ref{holon}) and
we have shown that under a re-sizing of the fiducial cell, the connection $c$ scales
according to equation (\ref{cscale}). Quantum corrections can arise
when $\mb c$ becomes on the order of one \cite{Willis:thesis} and thus we can generate arbitrarily
large quantum corrections by choosing a larger fiducial cell as long as $\mb$ is a
fixed constant. However, if $\mb$ scales as
\bq
	\mb \propto \frac{1}{\sqrt{|p|}}
\eq
then we find that the quantity $\mb c$ is invariant under a resizing of the
fiducial cell. A direct result of this in 
the $k=0$ model is that the bounce occurs
when the matter energy density is on the order of Planckian \cite{Ashtekar:BBII}
which is to be expected on physical grounds.
On the other hand, 
if $\mb$ is a fixed constant, the bounce can occur even at largely sub-Planckian
densities  and even a cosmological
constant can trigger a future recollapse of the universe \cite{Ashtekar:2006rx, Ashtekar:2006uz, Banerjee:2005ga}.
 The second motivation for this scaling comes
from the method proposed in \cite{Ashtekar:2003hd} to constrain the value
of $\mb$ based on using the minimum area eigenvalue of LQG in constructing
the Hamiltonian constraint operator. We will discuss this in more detail when we construct
the constraint operator. For now let us assume that $\mb$ is given as
\bq \label{mubar}
	\mb = \sqrt{\frac{\Delta}{|p|}}
\eq
where $\Delta$ is a constant to be fixed later.

The kinematical Hilbert space can then be constructed and a basis is given by eigenstates of the $\wh{p}$ operator
labeled by a real parameter $v$ with eigenvalues
\bq
	\wh{p} \kn = \frac{\kappa \g \hbar}{6} \Bigg( \frac{|v|}{K}\Bigg)^{2/3} \kn
\eq
where the constant $K$ is given by
\bq \label{K}
	K = \frac{2}{3} \sqrt{\frac{\kappa \g \hbar}{6 \Delta}} \,.
\eq
Similarly the states $\kn$ are eigenstates of the fiducial cell volume operator
\bq
	\wh{V}_{\text{cell}} \, \kn = \Big(\frac{\kappa \g \hbar}{6}\Big)^{2/3} \,  \frac{|v|}{K} \,\kn
\eq
The parameter $v$ runs over the entire real line, but the spectrum is discrete
in the sense that the states $\kn$ are normalizable satisfying
\bq
	\braket{v'}{v} = \delta_{v' v}
\eq
A general quantum state is a continuous sum over the basis states $\kn$ as well
as any matter degrees of freedom. We will interest ourselves in the inclusion
of a scalar field degree of freedom whence a general quantum state
 is given by
\bq
	\ket{\Psi} = \int\! d\phi \sum_{v} \; \Psi(v, \phi) \,\ket{v, \phi}
\eq
with the kinematical inner product between two states given by
\bq \label{KIP}
	\braket{\Psi_1}{\Psi_2}_{\text{kin}} = \int\! d\phi \sum_{v} \,\overline{\Psi}_1(v, \phi)\; \Psi_2(v, \phi)\,.
\eq
A quantum state  which lies in the kinematical Hilbert space has
finite kinematical norm which implies
\bq
	\int\! d\phi \sum_{v} \,\overline{\Psi}(v, \phi)\; \Psi(v, \phi) < \infty \,.
\eq

This constitutes the kinematical Hilbert space as well as the action
of the basic flux operator $\wh{p}$. Additional basic operator are required in
the form of holonomy operators which can be built using
the formula (\ref{holon}) and the basic exponential operators
\bq
	\wh{h}_{\pm} = \exp(\mp i \wh{\mb c}/2) \,.
\eq
The basis $\kn$ has been chosen such that the exponential operators
act simply as shift operators
\bq
	\wh{h}_{\pm} \Psi(v) = \Psi(v \pm 1) \,.
\eq
An important feature of the quantization is that since holonomies
form the basic configuration variables, there is no basic
operator corresponding to the connection $\wh{c}$. In order
to construct such an operator, one has to approximate it using 
the basic holonomy operators. An example of this is given
by the Hamiltonian constraint to which we turn now.

\subsection{Quantum Difference Equation}
The next step in quantization is to construct a Hamiltonian constraint
operator that is self-adjoint on the kinematical Hilbert space. The classical
expression for the gravitational part of the constraint is again given
by
\bq\label{CGR}
	\CGR = -  \frac{6 }{\g^2}  \sqrt{|p|} \,\Big(c^2 -  \g^2 \Om^{2/3} \Big)
\eq
which is equivalent to the $k=0$ model up to the $\g^2 \Om^{2/3}$ term in
the parentheses. The main complication in constructing the gravitational part
of the Hamiltonian constraint operator is the lack of an operator for the bare connection.
Thus the $c^2$ term must be quantized using holonomies. Following the results
from the $k=0$ model, the following classical re-expression 
\bq
	\CGR =  -\frac{4}{\kap \hbar \g^3 \mb^3} \sum_{ijk}\epsilon^{ijk} \; \tr \Big[\big(
	h_i h_j h_i^{-1} h_j^{-1} \non \\- 2 \mb^2  \g^2  \Om^{2/3}\tau_i \tau_j \big)
	h_k \left\{ h_k^{-1} , V \right\} \Big]
\eq
can be shown to give back the classical expression (\ref{CGR}) in the limit
as $\mb$ is taken to zero. This expression is now readily quantizable with
the major non-triviality being that we can not take the limit as $\mb$ goes
to zero as that would require a $\wh{c}$ operator. Thus in the quantum constraint
operator we do not take the limit, instead leaving $\mb$ to be a finite parameter given
by the expression (\ref{mubar}).

In order to constrain the parameter $\Delta$ in the definition of $\mb$ (or
equivalently the parameter $K$ in (\ref{K})), in the $k=0$ model
one can connect to the full theory of LQG by shrinking $\mb$ until the closed loop
spanned by the edges of the holonomies $h_i h_j h_i^{-1} h_j^{-1}$ has the minimum
physical area eigenvalue of LQG. 
This fixes the value of $\Delta$ to be equal to the minimum area eigenvalue
of LQG $\Delta = 2\sqrt{3} \pi \gamma \lp^2$ which implies
that  $K$ is given by 
$K = \frac{2 \sqrt{2}}{3 \sqrt{3 \sqrt{3}}} 
$\cite{Ashtekar:BBII}.
However, in the $k=-1$ model this interpretation does not hold since the edges
do not close. Thus we can not a priori make the same assignment of $K$.
We can however turn to the $k=+1$ model for guidance. There, the quantization
has been performed using holonomies of the extrinsic curvature where the loop similarly
does not close \cite{Bojowald:2003mc,Bojowald:2003xf}. The quantization
of the $k=+1$ involving holonomies of the connection and using a closed
loop for the constraint operator appears in \cite{Newkplus1} and there
the value of $K$ is constrained to the same value as the $k=0$ model using
the same procedure. Furthermore, the quantization using holonomies of the
connection is quantitatively similar to the one using holonomies of the extrinsic
curvature in the $v \gg 1$ regime which can be taken as evidence
that the same value of $K$ should be used in both quantizations.
With this in mind, we will leave this issue open and assume  that the parameter $K$ is
on the order of one without explicitly fixing its value.

With the caveats mentioned, the construction of the constraint operator follows that
of the $k=0$ model (see \cite{Ashtekar:BBII} for details). The action on the 
the operator is given by
\bq
	\wh{\C}_{GR} \Psi(v) = f_{+}(v) \,\Psi(v+1) + f_0(v)\, \Psi(v) \non \\
	+ f_{-}(v) \,\Psi(v - 1)
\eq
with the functions $f$ given by
\bq
	f_{+}(v) &=& \frac{27}{16} \sqrt{\frac{\kappa \g \hbar}{6}} \frac{K}{\g^2}
	\, \big| v+2\big| \; \bigg| \big| v+1 \big| -  \big| v+3 \big|  \bigg| \\
	f_{-}(v) &=&  f_{+}(v-4) \\
	f_{0}(v) &=& - f_{+}(v) - f_{-}(v) + g(v)
\eq
and the function $g(v)$ representing the modification coming from the $k=-1$ model
given explicitly as
\bq
	g(v) = \frac{3 \Om^{2/3}}{K^{1/3}} \sqrt{\frac{\kappa \g \hbar}{6}}
	\;|v|^{1/3} \,  \bigg| \big| v+1 \big| -  \big| v-1 \big|  \bigg|
\eq
Thus the contribution from the $k=-1$ model amounts to the addition
of a term $g(v)$ that acts diagonally on the basis states $\kn$.

To discuss dynamics and interpret the difference equation we can add
matter in the form of a massless scalar field as done in \cite{Ashtekar:BBII}. Since
the difference equation will be of similar form the the $k=0$ model most of the results
remain valid. With a massless scalar field, the full constraint is given
by
\bq
	\wh{C} = \wh{C}_{GR} + \kappa \wh{p}^{-3/2} \wh{P}_{\phi}^2
\eq
where $\Pp$ is the canonical momentum to the scalar field. 
Since the Hamiltonian is independent of the scalar field $\phi$,
the conjugate momentum $\Pp$ is a constant of motion classically.
The classical Friedmann equation is given by
\bq
	H^2 = \frac{\kappa}{6} \frac{\Pp^2}{\Om^{2} a^6} + \frac{1}{a^2}
\eq
which can be solved explicitly in terms of conformal time $d\eta = a dt$ giving
\bq
	a^2(\eta) = \sqrt{\frac{\kappa \Pp^2}{6 \Om^2}} \sinh(2\eta)
\eq
and similarly the scalar field evolves as
\bq
	\phi(\eta) = \half \sqrt{\frac{6}{\kappa}} \ln(\tanh\eta) + \phi_0
\eq
Both are monotonic functions and thus can play the role of emergent time. If we choose the
scalar field to play the role of emergent time the evolution of $a$ is given by
\bq
	a^2(\phi) = \sqrt{\frac{\kap \Pp}{6}}\; \text{csch}\sqrt{2\kap/3}(\phi-\phi_0) 
\eq

For the quantization of the matter part of the constraint, 
 the operator for $\Pp$ acts 
simply as $\wh{P}_{\phi} = -i \hbar\, \partial / \partial\phi$. The
inverse volume operator $\wh{p}^{-3/2}$ requires careful treatment
as the naive inverse of the $\wh{p}$ operator does not lead to a densely
defined self-adjoint operator owing to the fact the the {\em normalizable}
state $\ket{v=0}$ lies in the spectrum of the $\wh{p}$ operator.
Using techniques from the full theory \cite{Thiemann:1996aw,Thiemann:1997rt},
 the application to LQC leads to a bounded self-adjoint operator \cite{Bojowald:2002ny}
with eigenvalues given by \cite{Ashtekar:BBII}
\bq
	\wh{p}^{-3/2} \Psi(v) = \Big(\frac{6}{\kap \g \hbar}\Big)^{3/2} B(v) \Psi(v)
\eq
with the function $B(v)$ given by
\bq\label{B}
	B(v) = \Big(\frac{3}{2}\Big)^{3} K |v| \Big| |v+1|^{1/3} - |v-1|^{1/3}    \Big|^3
\eq
This  inverse volume operator represent one choice among many possible choices
of the types that have been explored in \cite{Bojowald:2002ny}. In particular there is freedom
to use a particular spin $J$ SU(2) representation to define the holonomies.
The operator shown here corresponds to using the fundamental representation
($J=1/2$) in accordance with arguments indicating that the theory should
be quantized using that value\cite{Vandersloot:2005kh,Perez:2005fn}.
The behavior of $B(v)$ changes for $v < 1$ and $v>1$.
For $v < 1$, $B(v)$ behaves polynomially and increases for large values
of $v$ while it vanishes at the singularity $v=0$. For $v > 1$, $B(v)$ approaches
the classical expression $B(v) \approx \Big(\frac{\kap \g \hbar}{6}\Big)^{3/2} p^{-3/2}$.
The physical meaning of $v < 1$ is dependent on the choice of fiducial cell and we
will discuss this in more detail in section \ref{Discussion}.

With the matter constraint operator, the difference equation can be rearranged into
the form
\bq
	\frac{\partial^2 \Psi(v, \phi)}{\partial\phi^2} =
	- \wh{\Theta}\Psi(v,\phi) \equiv  -( \wh{\Theta}_0 +\wh{\Theta}_{-1}) \Psi(v,\phi)
\eq
with the $\wh{\Theta}_0$ operator equivalent to the $\wh{\Theta}$ operator in the $k=0$ model \cite{Ashtekar:BBII}
\bq
	\wh{\Theta}_0 \Psi(v) &=&- B(v)^{-1} \Big[ C^{+}(v) \Psi(v+4) + C^{0}(v) \Psi(v) \non \\
	&&\;\;\;\;\;\;+ \;C^{-}(v) \Psi(v) \Big]\\
	C^{+}(v) &=& \frac{3 \kappa K}{64} \, \big| v+2\big| \; \bigg| \big| v+1 \big| -  \big| v+3 \big|   \bigg| \\
	C^{-}(v) &=& C^{+}(v-4) \\
	C^{0}(v) &=& - C^{+}(v) - C^{-}(v)\,.
\eq
The $k=-1$ model contributes the $\wh{\Theta}_{-1}$ operator which acts diagonally
on $\Psi(v)$ as
\bq
	\wh{\Theta}_{-1} \;\Psi(v)  = -B(v)^{-1} \frac{\kappa \g^2 \Om^{2/3}}{12 K^{1/3}} 
	\;|v|^{1/3} \non \\ \times \bigg| \big| v+1 \big| -  \big| v-1 \big|  \bigg| \;\Psi(v)
\eq 

The combined operator $\wh{\Theta}$ is self-adjoint\footnote{
Technically $\wh{\Theta}$ is self-adjoint on
the Hilbert space $L^2(\RBohr, B(v) d\mu_{\text{Bohr}}) $ where
$\RBohr$ refers to the almost periodic functions. The extra factor of $B(v)$ is
due to the fact the $\wh{C}_{GR}$ is self-adjoint in the kinematical Hilbert space $L^2(\RBohr, d\mu_{\text{Bohr}}) $  and that $\wh{\Theta} \propto
B^{-1}(v) \wh{C}_{GR}$}
but not positive definite. If we restrict ourselves to the positive part of
the spectrum of $\wh{\Theta}$ then the physical inner product can 
be constructed in a simple fashion. Namely we restrict to eigenstates
$e_{\omega}(v)$ of the $\wh{\Theta}$ operator: $\wh{\Theta} e_{\omega}(v) = \omega^2 e_k(v)$. Since $\wh{\Theta}$ is self-adjoint and
we are restricting to the positive part of the spectrum, by spectral analysis
we can construct an operator corresponding to the square root
 $\sqrt{\wh{\Theta}}$. Solutions to the difference equation then split
into positive and negative frequency solutions satisfying
a  
 first order Schrodinger like equation
\bq \label{diffeqs}
	\mp i \frac{\partial \Psi(v, \phi)}{\partial\phi} = \sqrt{\wh{\Theta} } \; \Psi(v, \phi)
\eq
In this form, the difference equation is like a standard evolution equation
in terms of the scalar field $\phi$.
We can restrict to the positive frequency solution space when considering
physical wavefunctions whence the physical inner product is given in
analogy with the Schrodinger inner product of quantum mechanics as
\bq
	\braket{\Psi_1}{\Psi_2}_{\text{phys}} = \sum_{v} B(v) \,\overline{\Psi}_1(v, \phi_0)\, \Psi_2(v, \phi_0) \,.
\eq
As in quantum mechanics, the physical inner product can be evaluated at
any `time' $\phi_0$, and the difference equation  (\ref{diffeqs})
guarantees that the result is independent of $\phi_0$.

Lastly to interpret the physical wavefunctions requires the evaluation of
expectation values of observables. Technically, we require Dirac observables
which correspond to quantum operators which commute with the constraint
operator so as to lead to unambiguous gauge invariant observables. Following
the $k=0$ model \cite{Ashtekar:BBII} the scalar field momentum 
$\wh{P}_{\phi} = -i \hbar\, \partial / \partial\phi$ is an observable whose operator
trivially commutes with the constraint operator. An additional observable
is the value of $v$ at a given instant in `time' $\phi_0$ labeled
$v|_{\phi_0}$. The expectation value of this observable is given by
\bq
	\braketfull{\Psi}{\wh{v}|_{\phi_0}}{\Psi} = \frac{\sum_{v} B(v) \, v \,\overline{\Psi}(v, \phi_0)\, \Psi(v, \phi_0) } {\braket{\Psi}{\Psi}_{\text{phys}}}
\eq

\section{Dynamics}

With the inclusion of the massless scalar field, the resulting dynamics
and interpretation of the theory
can be understood by constructing suitable semi-classical states \cite{Singh:2005xg,Ashtekar:2006uz,Ashtekar:BBII}.
The dynamics of the theory is most easily understood by choosing the scalar field $\phi$ to play the
role of the internal clock. Following the procedure set forth in \cite{Ashtekar:2006uz,Ashtekar:BBII},
eigenfunctions of the $\wh{\Theta}$ operator are calculated and then Fourier
transformed to get physical wavefunction solutions to the quantum difference
equation.  Thus given the eigenfunctions $e_{\omega}(v)$ we choose
a Gaussian profile $e^{-(\omega-\omega_*)^2/2\sigma^2}e^{i \omega \phi_*}$
peaked around a large value of the scalar field momentum $\Pp = \hbar \omega_*$
with spread $\sigma$ and peaked around a value of the scalar field $\phi_*$.
Physical wavefunctions are constructed throuh the Fourier transform
\bq \label{Ftrans}
	\Psi(v, \phi) = \int_{-\infty}^{\infty} \!\! d\omega \, e^{-(\omega-\omega_*)^2/2\sigma^2}e^{i \omega \phi_*} e_{\omega}(v)\, e^{i \omega \phi} 
\eq
which are thus by construction solutions to the difference equation.
Numerically, the procedure is to first calculate the eigenstates $e_{\omega}(v)$.
Typically, there is a two-fold degeneracy in the eigenstates, and
this is removed by choosing the eigenstate that matches the positive frequency
Wheeler-DeWitt solution\footnote{Wheeler-DeWitt solutions
are eigenstates of the operator $\underline{\wh{\Theta}}$ which is the
continuous differential operator that approximates the difference operator
$\wh{\Theta}$ in the large $v$ limit. See \cite{Ashtekar:2006uz,Ashtekar:BBII} for more details.} for large $v$. Once the eigenstates are calculated, the Fourier transform
(\ref{Ftrans}) is calculated using the  Fast Fourier transform
algorithm.

\begin{figure}[ht]
\begin{center}
\includegraphics[width=8cm, keepaspectratio]
        {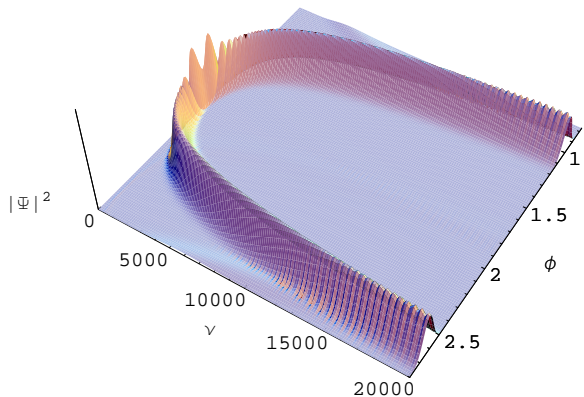}
\end{center}
\caption{Evolution of semi-classical state initially peaked at a large
value of $v$. The state remains sharply peaked and bounces before
reaching the singularity $v = 0$. After the bounce, the state continues to remain
sharply peaked and leads to an expanding universe. The values of
the numerical parameters used in the simulation were $\omega_* = 700$,
$\sigma = 20$, $\Om = 1$, and $K=1/2$.}
\label{3dplot}
\end{figure}

An example of a numerical simulation is shown in figure \ref{3dplot}. The state is initially peaked
around a large value of $v$ and evolves towards the singularity while remaining
sharply peaked. Instead of plunging into the singularity as expected from
the classical dynamics, the state bounces leading to an expanding universe. 
The results of the quantum dynamics are qualitatively similar to the $k=0,+1$ models
\cite{Ashtekar:BBII,Newkplus1} and the bounce occurs when the energy density
of the scalar field is Planckian.

The behavior of the dynamics can be understood in terms of an
effective classical description. This amounts to considering an
effective modified Hamiltonian constraint through which effective classical
equations of motion are calculated. Note that by nature this sort of effective
description can not completely encode the predictions from the quantum theory
and care must be taken when applying the effective theory in more general settings.
In particular if the wavefunction becomes non-sharply peaked, then additional modifications
to the dynamics are expected to become appreciable \cite{Bojowald:2006gr}. In the numerical
simulations performed for this work, the wavefunction remains sharply peaked throughout the 
evolution, and the effective description provides an accurate description which we show
explicitly now.

The effective Hamiltonian is given by (see \cite{Date:2004zd, Banerjee:2005ga, Vandersloot:2005kh, Willis:thesis, Bojowald:2006gr, Taveras2006} for various discussions
on the issue)
\bq
	\Ham_{eff} = -\frac{3 \sqrt{|p|}}{\kap \g^2 \mb^2} \, \sin^2(\mb c) +\frac{3 \sqrt{|p|} \Om^{2/3}}{\kap}+ |p|^{3/2}
	\rhom
\eq
where again $\mb$ is a function of $p$ given by
\bq
	\mb = \sqrt{\frac{\Delta}{|p|}}\, . 
\eq
Note that in this effective Hamiltonian, we are implicitly assuming
the $v \gg 1$ limit. In particular, in this limit the $B(v)$ eigenvalues
that would appear in the matter part of the Hamiltonian are approximated
by the classical expression; namely
\bq
	B(v) = \frac{K}{v} + \Or(v^{-3}) 
\eq
and thus the matter density takes on its classical form 
\bq \label{rhomeff}
	\rhom = \frac{\Pp^2}{2 p^3} + \Or(p^{-9/2})\,.
\eq
In this effective Hamiltonian we are therefore ignoring the inverse volume
corrections to the matter Hamiltonian and will show that this is
a good approximation by comparison with the quantum dynamics.

With this effective Hamiltonian we can derive an effective Friedmann equation. To do
this first we note that the left hand side of the Friedmann equation involving the Hubble
rate squared can be written as
\bq
	H^2 = \Big(\frac{\dot{a}}{a}\Big)^2 = \frac{1}{4}\Big(\frac{\dot{p}}{p}\Big)^2
\eq
which is a simple consequence from the fact that $p \propto a^2$ from equation (\ref{untild}).
The time derivative $\dot{p}$ is calculated from Hamilton's equation $\dot{p} = \{p, \Ham_{eff}\}$ giving
\bq
	H^2 = \frac{1}{4}\Big(\frac{\dot{p}}{p}\Big)^2 &=&  \frac{1}{\g^2 \mb^2 |p|} \sin^2\mb c \;\cos^2\mb c \non \\
	&=& \frac{1}{\g^2 \mb^2 |p|} \sin^2\mb c \; (1 - \sin^2\mb c) \,.
\eq
Lastly we can use the vanishing of the Hamiltonian to relate $\sin^2\mb c$ to $\rhom$ which gives
\bq
	\sin^2\mb c = \g^2 \mb^2 \Om^{2/3} + 
	\frac{\kap \g^2 \mb^2 |p|}{3} \rhom \,.
\eq
Putting these together and writing in terms of the scale factor $|p| = \Om^{2/3} a^2$
we get for the effective Friedmann equation
\bq \label{Friedeff}
	H^2 = \Big( \frac{\kap}{3} \rhom + \frac{1}{a^2}\Big) \Big(1 - \frac{\g^2 \Delta}{a^2} - 
		\frac{\kap \g^2 \Delta}{3} \rhom   \Big)
\eq
The first term in parentheses is the classical right-hand side of the Friedmann equation
and thus the second term in parentheses represents the quantum modifications. The bounce
can be understood as arising when second term vanishes; namely, when the matter density
reaches a maximum
\bq \label{rhomax}
	\rho_{max} = \frac{3}{\kap \g^2 \Delta} - \frac{3}{\kappa a^2}
\eq
where the first term is precisely the same form as the critical density $\rhoc= \frac{3}{\kap \g^2 \Delta}$ arising in the $k=0$ model
and the second term forms an additional contribution from the $k=-1$ model.
Notice that the actual value of the matter density at the bounce point, depends on the value
of the scale factor at the bounce point. To determine the bounce scale factor $a_c$ and the value
of the bounce energy density for the massless scalar field, we can solve for when the matter density
equals the maximum value
\bq
	\frac{\Pp^2}{2 \Om^2 a_c^6} = \frac{3}{\kap \g \Delta} - \frac{3}{\kappa a_c^2}
\eq 
If the scalar field momentum is sufficiently large, then $a_c$ is sufficiently large that the second
term is negligible and we find that the bounce energy density agrees with the form of the $k=0$ critical density
$\rho_{max} \approx \rhoc $. The actual value of $\rhoc$ is dependent
explicitly on the value of $\Delta$ which  by (\ref{K}) depends on the value of $K$.
If $K$ is on the order of one, then (\ref{K}) implies that $\Delta$ is on the order
of the Planck length squared, and one finds that $\rhoc$ is on the order
of the Planck density.  From the arguments of
\cite{Ashtekar:BBII}, the critical density in the $k=0$ model is valued at $\rhoc = .82 \rho_p$.

It is evident from the effective Friedmann equation (\ref{Friedeff}) and from
the form of the maximum energy density (\ref{rhomax}) that arbitrary matter
with positive energy density will trigger a bounce. Furthermore, the effective
Friedmann equation predicts a minimum scale factor $a_{min}$ that
the open universe can reach. Namely, even in the vacuum energy density
case the right hand side of the effective Friedmann equation is negative
and thus forbidden for values of the scale factor below
\bq
	a_{min} = \gamma \sqrt{\Delta}
\eq 
which again is on the order of the Planck length if $\Delta$ is on the order of
the Planck length squared. Thus the open model constructed here predicts
a vacuum repulsion in the high curvature regime.

\begin{figure}[ht]
\begin{center}
\includegraphics[width=8cm, keepaspectratio]
        {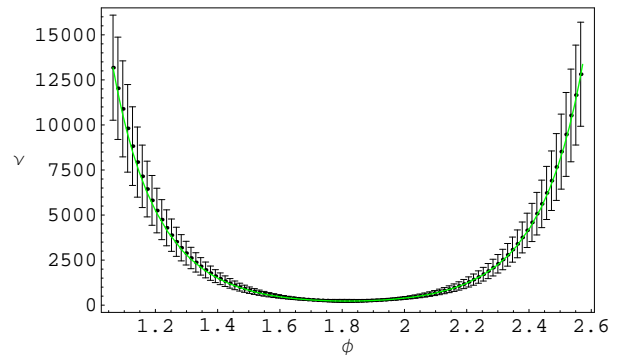}
\end{center}
\caption{Expectation value (dots) of $v|_{\phi}$ observable
with the error bars representing the dispersion. The expectation
values are approximated well by the predicted values (solid line) from
the effective Friedmann equation (\ref{Friedeff}).}
\label{vmeanplot}
\end{figure}

We can compare the predictions of the effective Friedmann equation
with the quantum dynamics as a method of testing the validity of
the effective theory.
In figure \ref{vmeanplot}, the expectation value of observable $<\wh{v}|_{\phi_0}>$ 
is plotted along with the spread $<\wh{\Delta v}|_{\phi_0}>$. The solid line is the trajectory predicted
from the effective Friedmann equation (\ref{Friedeff}) which agrees quite well with
the expectation values. We see that the effective Friedmann equation accounts
for the bounce at the right moment and agrees very well in the post bounce
regime. This testifies as to the validity of the effective theory in the 
massless scalar field model considered. Furthermore, we have ignored
the inverse volume corrections to the matter part of the effective Hamiltonian
and thus the quantum dynamics are not sensitive to those effects. The reason
for this is that the bounce occurs at a value of $v$ much larger than one. In particular
for the  values of the parameters chosen
in figure \ref{vmeanplot}, the bounce value of $v$ is 228.015. In order to probe
the small $v$ regime, one would need a semi-classical state with a small value
of $\Pp$ yet such states behave non semi-classically with large spread and thus
the effective description would not be valid and the quantum state would not
be a good description of our universe..

However, as we mentioned one should
keep in mind that in more complicated models, the effective theory shown
here can in
principle deviate from the quantum dynamics with deviations that may depend
of the quantum state. Thus it is an open issue to
understand better in what regimes the deviations occur and 
whether or not the deviations can be accounted for in a more complete
effective picture. An  effective theory that takes into
account the quantum degrees of freedom (such as the spread of the wavefunction)
can be found in \cite{Bojowald:2006gr}, and thus merits testing with the
quantum dynamics in more complicated scenarios.

\section{Discussion}\label{Discussion}

We have shown explicitly that a successful loop quantization of the $k=-1$ model
exists with the correct semi-classical limit. In this quantization the results
of the $k=0,+1$ models are extended and the classical singularity can be resolved
even leading to a big-bounce with a massless scalar field. This is further testament
to robustness of the predictions of LQC.

Several caveats of the model require discussion. First is that our model was constructed
using holonomies of the extrinsic curvature as opposed to holonomies of the connection
as done in the full theory. The reason for using this  quantization
is that the holonomies of the connection (an example of which
is given in formula \ref{hA}) are not almost periodic functions thus
rendering a loop quantization difficult.
As stated, this technique has been utilized in the 
loop quantization of the spherically symmetric models as well as in the inhomogeneous
cosmological model of \cite{Bojowald:2006qu}. An important question is therefore
what are the implications of the quantization using holonomies of the extrinsic curvature.

We can turn to the closed 
 $k=+1$ model where both quantizations have been performed, with 
holonomies of the extrinsic curvature being used in the earlier work \cite{Bojowald:2003mc} while
holonomies of the connection comprising the quantization in the more recent work of \cite{Newkplus1}. The two quantizations can be shown to be in agreement in
the $v \gg 1$ limit, with the differences restricted to the small volume $v \ll 1$ regime.
We can
understand the reason for this behavior in the following heuristic way. The holonomies of the connection
consist of exponentials of $\mb$ times the connection; i.e., $\mb (\g K_a^i + \G_a^i)$
where for the non flat models the spin connection components $\G_a^i$ are constant
valued.
Since we are taking $\mb$ to scale as $p^{-1/2}$ (equivalently $v^{-1/3}$),
then for large values of $v$ the quantity $\mb \G_a^i$ is guaranteed to be small.
Thus, the difference of holonomies of the connection and extrinsic curvature are expected
to be negligible in the large $v$ limit. This is precisely what is observed in the $k=+1$ model.

Therefore, the results of the $k=+1$ model indicate that for the $k=-1$ quantization
presented here, the results are expected to be valid in the $v \gg 1$ regime. This
does not affect any of the results presented here provided that the semi-classical
state does not approach the $v < 1 $ regime. As we have mentioned,
the bounce occurs at a value of $v \gg 1$ for universes which behave semi-classically.
Furthermore, in the $k=+1$ model similarly the bounce occurs at $v \gg 1$
for universes which reach macroscopic size before recollapsing \cite{Newkplus1}.
Thus we expect that the physical results presented in this paper, such as the quantum bounce, are largely insensitive
to whether the quantization is performed using holonomies of the connection or
extrinsic curvature.

Additionally there is the issue of the dependence of the quantum results
on the size of the fiducial cell. First we can ask if the effective
Friedmann equation (\ref{Friedeff}) is dependent on the fiducial
cell and therefore the prediction of the quantum bounce.
Classical quantities such as the scale factor $a$ and matter energy density
do not make reference to the fiducial cell and thus do not rescale. This
implies that the effective Friedmann equation (\ref{Friedeff}) is invariant
under a change in fiducial cell. Note that the result crucially depends
on the fact that $\mb$ is not taken to be a constant, but scales
as $p^{-1/2}$. Thus the prediction of the bounce does not make
reference to the fiducial cell.

The same statement can not be made about the inverse volume corrections
appearing in the quantum matter density of the scalar field. The eigenvalues $B(v)$
give back the classical behavior for $v \ll 1$ but in general  behave as
\bq
	B(v) \propto \left\{  
		\begin{array}{lcl}
			v^4  &&v \ll 1 \\
			v^{-1} &&v \gg 1 
		\end{array}
	\right.
\eq
The parameter $v$ is proportional to the physical volume of the
fiducial cell, and thus must scale if we resize the fiducial cell.
The exact scaling under a resizing of the fiducial cell
$\Om \ra \Om' = \xi^3 \Om$, is given as
\bq
	v \ra v' = \xi^3 v
\eq
For a given value of the scale factor, a larger fiducial cell implies a larger value
of $v$. 
In terms of the scale factor, $v$ is related as
\bq
	v 	= \Om K \; \Big( \frac{6}{8\pi \g}\Big)^{3/2}  \frac{a^3}{\lp^3}
\eq
which makes evident that the value of $v$ depends explicitly on
the fiducial cell volume $\Om$ for a {\em fixed} value of the scale factor.
If we enlarge the fiducial cell, then the value of $v$ should also
increase which in turn {\em reduces} the effects of the inverse volume
eigenvalues. Vice versa, a smaller fiducial cell implies stronger
inverse volume effects.

Thus, when considering phenomenological applications involving
the inverse volume modifications, one must specify
the scale at which the inverse volume effects are non-negligible. In
other words, the critical scale separating the quantum regime from
the classical regime corresponds to $v = 1$ which in terms
of a critical scale factor $a_*$ gives
\bq
	a_* = \sqrt{\frac{8\pi \g}{6}} \frac{\lp}{K^{1/3}} \; \Om^{-1/3}
\eq
which indicates the explicit dependence on the fiducial cell. Again, a larger
value of $\Om$ implies a smaller $a_*$ which pushes the quantum effects
into the higher curvature regime and vice versa. If the fiducial cell
volume $\Om$ and $K$ are on the order of one, then $a_*$ is on the order
of the Planck length, but note that the  critical scale is  not necessarily
Planckian.

The issue of the scale dependence of the inverse volume modifications
occurs additionally in the $k=0$ model where again a fiducial cell is required
to quantize the spatially infinite model (see discussions in \cite{kevthesis,Ashtekar:BBII}). The preceding arguments remain valid
for this model and a scale must be introduced. On the other hand, the compact $k=+1$
model does not require a fiducial cell since the spatial integrations do not diverge.
There, inverse volume modifications occur when the physical volume of the entire
universe is Planckian. In other words, the scale at which the quantum effects occur
is provided by physical volume of the universe. For the closed model this
is equivalent to the high curvature Planckian regime.

Since the scale at which the inverse volume effects occur is given by
the physical size of the fiducial cell in the $k=0,-1$ models, an important
issue is to determine what sets the scale in loop quantum cosmology.
The fiducial cell was introduced in order to regulate the infinite spatial
integrations appearing in the action and Hamiltonian and thus
is not expected to be physically relevant. One possibility
is that the scale is provided in an {\em inhomogeneous} treatment
of loop quantum cosmology. An inhomogeneous model of loop quantum
cosmology has been developed in \cite{Bojowald:2006qu} based
on a fixed lattice quantization. In that model, the scale corresponds
to the physical size of the lattice links. Yet, the inhomogeneous
model does not provide a prescription to determine the size of the scale, which
must be specified by hand and is not necessarily tied to matter degrees of freedom or the curvature scale. The naive expectation would be that
the lattice spacing should be Planckian in size, but if the model describes
the current universe then we would expect to see
 inverse volume modifications occurring today, a prediction which
is clearly ruled out by observations.

Whatever determines the scale inherent in LQC models, one is faced with
constraining the predictions with observations. As mentioned, 
if the scale is too small, then inverse volume corrections
might be predicted in the near past which would alter the Friedmann dynamics and
be observationally detectable. If the lattice links of an inhomogeneous model provide the scale, the links must
be sufficiently larger than the Planck scale in the recent history of the universe, but presumably
not too large to spoil particle physics. If the scale provided by the lattice links
expands with the growing universe (i.e. the lattice links grow with the universe), then ensuring that they are not too large today,
while being not too small in the earlier universe could be challenging and might require
fine tuning. The inhomogeneous model of 
of \cite{Bojowald:2006qu} has the behavior that the lattice links expand with
the universe and thus would face this constraint.
However, as mentioned in \cite{Bojowald:2006qu}, one possibility is that
in a more systematically derived inhomogeneous lattice model of loop quantum cosmology, the scale provided by the lattice
links would dynamically change and thus might not grow with the expanding universe.
This type of behavior is mimicked in the homogeneous setting
when $\mb$ scales as a function of $p^{-1/2}$, a quantization feature which was
first proposed in \cite{Ashtekar:BBII} and has been utilized in
this paper. With this scaling behavior, the holonomy edges defining the Hamiltonian
constraint operator decrease in physical length with the expanding universe.
The results of the improved quantization appear better grounded on a
physical basis and thus this behavior would seem to be a requirement
for constructing inhomogeneous models. 

Furthermore,
we have shown that the inverse volume modifications play no important role
in the quantum dynamics for universes which behave semi-classically
since the bounce occurs for $v \gg 1$. Additionally, in the $k=+1$ model, for universes which grow to macroscopic size, again the inverse
volume modifications play no role \cite{Newkplus1}. We these indications,
along with the arguments that the ambiguity parameter $j$ should be
its lowest value $1/2$ \cite{Vandersloot:2005kh,Perez:2005fn},
these results give evidence  that the inverse volume modifications may not play a significant role
in the evolution of the universe.

\acknowledgments{The author would like to thank Abhay Ashtekar, Martin Bojowald, and Parampreet Singh for discussions, and to Abhay Ashtekar, Parampreet Singh, and 
Tomasz Pawlowski for
helpful comments on the manuscript. This work was supported in part by
the NSF grant PHY-0456913, the Eberly research funds of Penn State,
and the Marie Curie Incoming International Fellowship MIF1-CT-2006-022239.}


\end{document}